\documentclass[preprint,aps,aip]{revtex4-1}
\usepackage{amsmath}
\usepackage{color}
\usepackage{graphicx}
\usepackage{epsfig}

\begin{document}

\title{A general formalism of two-dimensional lattice potential on beam 
transverse plane for studying channeling radiation}

\author{Jack J. Shi and Wade Rush} 

\address{Department of Physics \& Astronomy, 
         The University of Kansas, Lawrence, KS 66045}

\begin{abstract}
To study channeling radiation produced by an ultra-relativistic electron beam channeling 
through a single crystal, a lattice potential of the crystal is required for solving the 
transverse motion of beam electrons under the influence of the crystal lattice. In this 
paper, we present a general formalism for this two-dimensional lattice potential of a 
crystal with a Lorentz contraction in the beam channeling direction. With this formalism, 
the lattice potential can be calculated without approximation from any given model of 
electron-ion interaction for an ultra-relativistic beam channeling in any crystal 
direction. The formalism presented should be the standard recipe of the lattice potential 
for studying the channeling radiation.

\end{abstract}


\maketitle

\section{Introduction}

There has been a renewed interest recently in channeling radiation \cite{Kuma} produced 
from an electron beam interacting with a crystal lattice for its potential application 
in hard X-ray production 
\cite{Piot13,Sen14,Wade,Azad06,Azad08,Azad11,Azad13,Gary90,Gary91,Genz90,Genz96}. 
When an ultra-relativistic electron enters a single crystal, rather than having random 
scattering, the electron will channel through the crystal lattice if its incident angle 
relative to a specific lattice orientation of the crystal is sufficiently small 
\cite{Lindh65}. During the channeling, the non-relativistic motion of the electron in 
the transverse plane that is perpendicular to the channeling direction could be strongly 
perturbed by the crystal lattice while the ultra-relativistic motion of the electron 
along the channeling direction is unperturbed. A high-intensity ultra-relativistic 
electron beam could produce high-brightness hard X-rays due to the perturbation of the 
transverse motion of beam electrons in a crystal. 

The channeling radiation is considered to be from the transitions between bounded Bloch 
eigenstates for the transverse motion of beam electrons during the channeling. To study 
the radiation theoretically and numerically, the interaction between the crystal lattice 
and beam electrons has been modeled in two different approaches, the one-dimensional (1D) 
planar \cite{Ander81,Ander85} and the two-dimensional (2D) axial \cite{Ander82,Ander85} 
channeling model. In the planar channeling model, the transverse motion of the electrons 
is assumed to be aligned in a single crystal direction during the channeling and the 
Bloch eigenstates for the transverse motion are solved from a 1D Schr\"{o}dinger 
equation. This 1D approximation is valid only if the coupling between the original 2D 
motion of the electrons in the transverse plane is negligible. Note that the lattice 
potential for the transverse motion of the channeling electrons is the result of a 
Lorentz contraction of the three-dimensional (3D) crystal lattice in the beam rest frame 
where the crystal travels with near the speed of light along (opposite to) the beam 
channeling direction and the 3D crystal lattice is pancaked into a 2D lattice on the beam 
transverse plane. The resulting lattice potential is strongly coupled in the 2D transverse 
plane and the transverse motion of a channeling electron in a crystal cannot be decoupled 
into 1D motion if the lattice potential dominates the Hamiltonian. At the bounded Bloch 
eigenstates for the channeling radiation, the transverse energy is negative (inside 
lattice potential wells) and the lattice potential dominates the transverse kinetic 
energy. Therefore, the 1D approximation of the planar channeling model cannot be 
justified rigorously and the study of the channeling radiation should be based on the 
Bloch eigenstates for the 2D transverse motion of beam electrons in a crystal. In order 
to solve the Bloch eigenstates in the 2D transverse plane, a lattice potential in the 
transverse plane is required. To obtain this potential of the 2D lattice, in the axial 
channeling model, the ions in the original crystal are first grouped into strings of the
ions along the channeling direction and the lattice potential is calculated by summing 
up the contributions of the strings of the ions \cite{Ander82,Ander85,Chouf99}. In this 
direct calculation of the lattice potential, the identification of the strings of the 
ions and the summation over the strings are cumbersome and, in some cases, difficult 
due to a complicated geometric relationship among lattice ions along the channeling 
direction. Moreover, the calculation with the strings of the ions has to be specifically 
tailored for each specific crystal structure and channeling direction, which can hardly 
be applied to other cases because of the difference in the strings of the ions with 
different crystal structure and different channeling direction. 

An accurate lattice potential on the transverse plane is necessary for calculating the 
Bloch eigenstates of the 2D transverse motion of beam electrons in a crystal. With a 
given interaction between a beam electron and individual ion in a crystal, such as the 
Born approximation for electron scattering from an ion with Doyle-Turner fitting 
parameters \cite{Doyle}, the question is what is the best approach, in terms of accuracy, 
mathematical simplicity, and general applicability to any crystal structure and 
channeling direction, for the construction of the lattice potential on the 2D transverse 
plane. In this paper, we present a new and better approach for this 2D lattice potential 
calculation. In this new approach, a lattice potential of the original three-dimensional 
crystal is first calculated from a given electron-ion interaction potential in the 
original unit cell coordinate of the crystal by taking the advantage of the native 
periodicity of the crystal lattice. This potential of the three-dimensional lattice in 
the unit-cell coordinate is then transferred into the beam coordinate that is aligned 
with the beam longitudinal (channeling) and transverse directions using rotational 
coordinate transformations. Lastly, the Lorentz contraction of the lattice can be easily 
accomplished mathematically by averaging the three-dimensional lattice potential in the 
beam coordinate along the channeling direction. In this approach, the calculation of 
the lattice potential is mathematically clean, systematic, and can be easily applied to 
any crystal with any channeling direction. This generic formulation of the 2D lattice 
potential is developed in Section II, Section III presents several examples of the 
channeling in different crystal direction, and Section IV contains a final remarks. 


\section{Formulation of Lattice Potential in Beam Transverse Plane}

When an ultra-relativistic electron beam channels through a crystal lattice with a 
specific orientation aligned along the beam longitudinal direction, the non-relativistic 
transverse motion of the electrons is perturbed by the lattice of ions in the crystal. 
To formulate this lattice potential, we consider an orthorhombic crystal lattice with 
each unit cell of the lattice containing $N$ ions. Let $\vec{r}_j$ with $j=1,\cdots,N$ 
be the local coordinates of the ions in a unit cell, where the origin of the local 
coordinate is at a corner of a unit cell, and $\vec{r}_{m}=(m_1a_1,\,m_2a_2,\,m_3a_3)$ be 
the global coordinate of a unit cell, where $(a_1,a_2,a_3)$ are lattice constants of the 
crystal and $\vec{m}=(m_1,m_2,m_3)$ are an integer vector. Let $\vec{X}=(X_1,X_2,X_3)$ 
be a global coordinate, referred as lattice coordinate, with the axes of $X_1$, $X_2$, 
and $X_3$ aligned with the primary crystal axes along [100], [010], and [001] crystal 
direction, respectively. When an electron travels through the crystal, the interaction 
between the electron and the crystal lattice can be calculated by a superposition of the 
interactions between the electron and each individual ion as
\begin{equation}
\label{v3D}
V_{3D}(\vec{X}\,) = \sum_{\vec{m}=-\infty}^{\infty}\sum_{j=1}^N 
                    V_{ion}(\vec{X} -\vec{r}_m - \vec{r}_j) 
\end{equation}
where $V_{ion}(\vec{r}\,)$ is the interaction potential between an electron and a single 
ion in the crystal. Note that $V_{3D}(\vec{X}\,)$ is periodic in the crystal, $i.e.$ 
$V_{3D}(\vec{X}+\vec{r}_l)=V_{3D}(\vec{X}\,)$, where $\vec{r}_l=(l_1a_1,l_2a_2,l_3a_3)$ 
and $(l_1,l_2,l_3)$ is any combination of integers. With the lattice native periodicity
$(a_1,a_2,a_3)$, $V_{3D}(\vec{X}\,)$ can be rewritten into a Fourier expansion of
\begin{equation}
\label{V(r)Exp}
V_{3D}(\vec{X}\,)=\sum_{\vec{k}=-\infty}^{\infty} V_{\vec{k}}\,e^{i\vec{G}\cdot\vec{X}}
\end{equation}
where $\vec{G}=2\pi(k_1/a_1,k_2/a_2,k_3/a_3)$ is the reciprocal lattice vector of the 
crystal. The Fourier expansion coefficient $V_{\vec{k}}$ can be calculated as
\begin{eqnarray}
\label{Vjack}
V_{\vec{k}} &=& \frac{1}{a_1a_2a_3} \sum_{\vec{m}=-\infty}^{\infty}\sum_{j=1}^{N}     
                \int\limits_{0}^{a_1}\int\limits_{0}^{a_2}\int\limits_{0}^{a_3}
                 V_{ion}(\vec{X} -\vec{r}_m - \vec{r}_j) 
                \,e^{-i\vec{G}\cdot\vec{X}}\,d\vec{X} \nonumber\\
&=&  \frac{1}{a_1a_2a_3} \sum_{j=1}^{N} \int\limits_{-\infty}^{\infty}
 \int\limits_{-\infty}^{\infty}\int\limits_{-\infty}^{\infty}
 V_{ion}(\vec{\xi}-\vec{r}_j) \,e^{-i\vec{G}\cdot\vec{\xi}}\,d\vec{\xi} \nonumber\\
&=& -\frac{2\pi\hbar^2}{m_e v_0}\,
     f_{ion}(\vec{G}\,) \,\sum_{j=1}^{N} e^{-i\vec{G}\cdot\vec{r}_j}
\end{eqnarray}
where $v_0=a_1a_2a_3$ is the unit-cell volume of the crystal, $m_e$ is the electron rest 
mass, and
\begin{equation}
\label{formfact}
f_{ion}(\vec{G}\,) = -\frac{m_e}{2\pi\hbar^2} \int\limits_{-\infty}^{\infty}
      \int\limits_{-\infty}^{\infty}\int\limits_{-\infty}^{\infty}
      V_{ion}(\vec{\xi}\,) \,e^{-i\vec{G}\cdot\vec{\xi}}\,d\vec{\xi} 
\end{equation}
is the atomic form factor of a single ion in the crystal \cite{Kittel}. Note that the 
lattice potential calculated from Eqs. (\ref{Vjack}) and (\ref{formfact}) includes exactly 
all the contributions of ions in a crystal and only approximation involved is in the 
modeling of the electron-ion interaction $V_{ion}$. With the Doyle-Turner approximation 
of the form factor \cite{Doyle}, the electron-ion interaction potential can be expressed 
as \cite{Ander85}
\begin{equation}
\label{3dt}
 V_{ion}(\vec{\xi}\,) = -\frac{16 \pi \hbar^2}{m_e}
    \sum_{i=1}^{M}\frac{\alpha_i}{(\beta_i/\pi)^{3/2}}\,e^{-4\pi^2|\vec{\xi}|^2/\beta_i} 
\end{equation} 
where $\alpha_i$ and $\beta_i$ with $i=1,\cdots,M$ are the parameters for fitting
$f_{ion}(\vec{G}\,)$ to the relativistic Hartree-Fock calculation of the form factor at 
values of $|\vec{G}|$ \cite{Doyle}. In the original Doyle-Turner's calculation, four 
Gaussian functions ($M=4$) were used in Eq. (\ref{3dt}). To improve the accuracy at 
relatively large value of $|\vec{G}|$, Chouffani and \"{U}berall used six Gaussian 
functions ($M=6$) for the fitting. The obtained values of $\alpha_i$ and $\beta_i$ for 
diamond, silicon and germanium were given in Ref. \cite{Chouf99}. With the electron-ion 
interaction potential in Eq. (\ref{3dt}), the Fourier transformation of the lattice 
potential of a three-dimensional crystal can then be calculated from Eqs. (\ref{Vjack}) 
and (\ref{formfact}) as
\begin{equation}
\label{Vk_3dPot}
V_{\vec{k}} =  -\frac{2\pi \hbar^2}{m_ev_0} 
   \sum_{i=1}^{M}\alpha_i\,e^{-\lambda_i^2|\vec{k}|^2/(2a)^2}\,
   \sum_{j=1}^{N} \,e^{-i2\pi(\vec{k}\cdot\vec{r}_j)/a}
\end{equation}
where $\lambda_i^2=\beta_i$ for $V_{ion}$ given in Eq. (\ref{3dt}) and
$\lambda_i^2=\beta_i+8\pi^2\left<\mu^2\right>$ for including the effect of thermal 
vibrations of the lattice with $\sqrt{\left<\mu^2\right>}$ being the root-mean-square 
displacement of the thermal vibration of the lattice \cite{Ander85}.

The lattice potential $V_{3D}(\vec{X}\,)$ in Eq. (\ref{V(r)Exp}) is expressed in the 
lattice coordinate that is aligned with the primary crystal axes but not aligned with 
the beam channeling direction. In order to study channeling radiation that is generated 
from the perturbation of the transverse motion of beam electrons under the influence 
of the lattice potential, the lattice potential in the beam transverse plane is needed. 
Since the beam is ultra-relativistic, the interaction between a beam electron channeling 
through the crystal and the ions in the crystal is too weak to have a significant effect 
on the longitudinal motion of the electron. The dependence of the lattice potential on 
the longitudinal coordinate of the electron can therefore be neglected by averaging the 
potential over the longitudinal direction. It is thus necessary to transform the lattice 
potential into a beam coordinate that is aligned with the beam channeling direction. 
Physically, this is just rotating the crystal lattice so that the crystal axis for the 
channeling is aligned with the beam direction. Let $\vec{r}=(x,y,z)$ be the beam 
coordinate where $z$ is along the beam channeling (longitudinal) direction of the beam and 
$(x,y)$ are two orthogonal coordinates in the beam transverse plane. The transformation 
from the lattice coordinate $\vec{X}$ to the beam coordinate $\vec{r}$ can be accomplished 
by a $3\times 3$ rotational matrix ${\bf R}$, $\vec{r}= {\bf R}\vec{X}$. For a beam 
channeling along the $[hkl]$ crystal direction, where three integers $h$, $k$, and $l$ 
are the Miller indices of crystal lattices \cite{Kittel}, ${\bf R}$ can be in general 
constructed with three consecutive transformations (see Fig. 1 for an illustration). 
The first transformation rotates the lattice coordinate $(X_1,X_2,X_3)$ in the $X_1-X_2$ 
plane by an angle of $\theta_1=\arctan(k/h)$ and the transformed coordinate is labeled 
as $(X'_1,X'_2,X_3)$. The axes of $X'_1$ and $X'_2$ are aligned with the $[hk0]$ and 
$[\bar{k}h0]$ crystal direction, respectively (see Fig. 1b). The second transformation 
is rotating $(X'_1,X'_2,X_3)$ in the $X'_1-X_3$ plane by an angle of 
$\theta_2=\arctan(l/\sqrt{h^2+k^2})$ and the transformed coordinate is labeled as 
$(X^{''}_1,X'_2,X'_3)$. The axes of $X^{''}_1$, $X'_2$, and $X'_3$ are aligned with 
$[hkl]$, $[\bar{k}h0]$, and $[h_yk_yl_y]$ direction, respectively, (see Fig. 1c) where 
\begin{equation}
(h_y\,,\,k_y\,,\,l_y) = (h\,,\,k\,,\,l)\times(-k\,,h\,,0) =
(-hl\,,\,-kl\,,\,h^2+k^2) 
\end{equation}
The third transformation is to switch the coordinate axes such that $x=X'_2$, $y=X'_3$, 
and $z=X^{''}_1$, where $z$ along the $[hkl]$ direction is the beam longitudinal 
coordinate (channeling direction) and $x$ along the $[\bar{k}h0]$ direction and $y$ 
along the $[h_yk_yl_y]$ direction are two orthogonal coordinates in the beam transverse 
plane, respectively. These three coordinate transformations can be expressed as 
\begin{equation}
\label{Rotation}
\vec{X}={\bf R}^{-1} \vec{r} =
 \left( \begin{array}{ccc}
\cos\theta_1 & -\sin\theta_1 &   0\\
\sin\theta_1 & ~\cos\theta_1 &  0 \\
0  & 0 &1\end{array} \right)
\left( \begin{array}{ccc}
\cos\theta_2 & 0  & -\sin\theta_2  \\
0      & ~ 1 &  0 \\
\sin\theta_2 & 0 & \cos\theta_2 \end{array} \right)
\left( \begin{array}{ccc}
\;0\;\, & \;\,0 \;\,& \;\,1\;  \\
1 & 0 & 0 \\
0 & 1 & 0 \end{array} \right) 
\left(\begin{array}{c} x \\ y \\ z \end{array}\right)
\end{equation} 
With the transformation matrix ${\bf R}$ in Eq. (\ref{Rotation}), 3D lattice potential 
$V_{3D}(\vec{X}\,)$ in Eq. (\ref{V(r)Exp}) can be expressed in the beam coordinate as 
\begin{equation}
\label{V3dRotate}
V_{3D}({\bf R}^{-1}\vec{r}\,) = \sum_{\vec{k}=-\infty}^{\infty} V_{\vec{k}}\,
         \exp\left[i(\vec{G}^T{\bf R}^{-1})\cdot\vec{r}\,\right] 
\end{equation}
where $\vec{G}^T=2\pi(k_1/a_1,k_2/a_2,k_3/a_3)$ denotes the row vector of $\vec{G}$ 
for the matrix multiplication. Since the crystal lattice has also periodicities 
along the orientations of the beam coordinate, $V_{3D}$ is periodic in $\vec{r}$. Let 
$(b_1,b_2,b_3)$ be the periodicities of the lattice along the $(x,y,z)$ direction, 
respectively. The periodicity of $V_{3D}(\vec{r}\,)$ requires
\[e^{ib_1(\vec{G}^T{\bf R}^{-1})_1}= 
  e^{ib_2(\vec{G}^T{\bf R}^{-1})_2}= 
  e^{ib_3(\vec{G}^T{\bf R}^{-1})_3}= 1\]
which yields a transformation of the reciprocal lattice vector 
\begin{equation}
\label{Grot}
 \left(\,\frac{k_1}{a_1}\,,\,\frac{k_2}{a_2}\,,\,\frac{k_3}{a_3}\,\right){\bf R}^{-1}
=\left(\,\frac{n_1}{b_1}\,,\,\frac{n_2}{b_2}\,,\,\frac{n_3}{b_3}\,\right)
\end{equation}
where $(n_1,n_2,n_3)$ are integers. The relationships between $(b_1,b_2,b_3)$ and 
$(a_1,a_2,a_3)$ and between $(n_1,n_2,n_3)$ and $(k_1,k_2,k_3)$ can be obtained from 
Eq. (\ref{Grot}) with the condition that both $(n_1,n_2,n_3)$ and $(k_1,k_2,k_3)$ are 
integers for a set of smallest values of $(b_1,b_2,b_3)$. Since the transformation in 
Eq. (\ref{V3dRotate}) is from the expansion in $\vec{X}$ with $(k_1,k_2,k_3)$ as the 
indices to an expansion in $\vec{r}$ with $(n_1,n_2,n_3)$ as the indices, $(k_1,k_2,k_3)$ 
as functions of $(n_1,n_2,n_3)$ are needed here. For each given set of $(n_1,n_2,n_3)$ in 
integers, therefore, $(k_1,k_2,k_3)$ have to be solved from Eq. (\ref{Grot}) as integers. 
With the coordinate of the 3D lattice potential of a crystal correctly aligned with the 
beam directions, the average of the lattice potential along the beam channeling ($z$) 
direction can easily be calculated as 
\begin{equation}
\label{avg}
V_{2D}(x,\,y) = \frac{1}{b_3} \int_{0}^{b_3}V_{3D}({\bf R}^{-1}\vec{r}\,)\,dz 
         = \sum_{\vec{k}=-\infty}^{\infty} V_{\vec{k}}\,
              e^{i2\pi(n_1x/b_1+n_2y/b_2)} \, \delta_{n_30} 
\end{equation}        
where $\delta_{n_30}$ is the Kronecker delta for $n_3=0$. The Fourier expansion of 
this projected lattice potential on the transverse plane can then be written as
\begin{equation}
\label{V(xy)}
V_{2D}(x,y)=\sum_{n_1=-\infty}^{\infty} \sum_{n_2=-\infty}^{\infty} 
            V_{n_1n_2}\,e^{i2\pi(n_1x/b_1+n_2y/b_2)}
\end{equation}
and the expansion coefficient can be obtained from $V_{\vec{k}}$ of the 3D lattice
potential that is calculated using Eqs. (\ref{Vjack}) and (\ref{formfact}) as
\begin{equation}
\label{V2dFourier}
V_{n_1n_2} = \sum_{\vec{k}=-\infty}^{\infty} V_{\vec{k}}\; 
             \delta\left(\vec{k}-\vec{k}_0(n_1,n_2)\right)
\end{equation}
where $\vec{k}_0(n_1,n_2)$ is the solution of $\vec{k}=(k_1,k_2,k_3)$ as functions of 
$(n_1,n_2,n_3)$ solved from Eq. (\ref{Grot}) with $n_3=0$ and the delta functions are 
the Kronecker delta for $\vec{k}=\vec{k}_0(n_1,n_2)$. 

It should be noted that the minima (potential wells) of $V_{2D}(x,y)$ in the transverse 
plane form periodically a 2D lattice that results from the projection of the original 
crystal lattice onto the transverse plane. The Lorentz contraction along the channeling 
direction of a three-dimensional lattice can create a more condense structure of ions in 
the transverse plane. The periodicity of $V_{2D}(x,y)$ could be smaller than $b_1$ and 
$b_2$ due to the averaging of 3D lattice potential $V_{3D}$ along the beam channeling 
direction. Mathematically, the change of the periodicity occurs when the summation over 
the ions in a unit cell in Eq. (\ref{Vk_3dPot}) zeros periodically for certain combinations 
of $(k_1,k_2,k_3)$ with $n_3(\vec{k}\,)=0$. If that occurs, the summation for the Fourier 
expansion of $V_{2D}(x,y)$ in Eq. (\ref{V(xy)}) can be rearranged with the correct values 
of the periods. To solve the Bloch eigenstates for the transverse motion of beam electrons, 
moreover, one needs to identify the primitive unit cell of the 2D lattice of $V_{2D}(x,y)$ 
and the $x$ and $y$ coordinate should be aligned with the axes of the primitive unit cell 
\cite{Kittel}. The crystal axes of the primitive unit cell and the lattice constants that 
are the periods of $V_{2D}(x,y)$ can be easily identified by examining the contour plot 
of $V_{2D}(x,y)$. If the coordinate $(x,y)$ in Eq. (\ref{V(xy)}) is not aligned with the 
primitive unit cell, an additional rotational transformation in the transverse plane is 
needed for $V_{2D}(x,y)$. 

\section{Examples}

To illustrate this general method for the lattice potential in the beam transverse plane, 
the potential is calculated for the channeling along [001], [110], [111], and [210]  
direction of a cubic crystal with a diamond-like atomic structure that has eight ions 
in a unit cell (e.g. diamond, silicon, germanium). The coordinate of the ions in a 
unit cell is listed in Table I and the lattice constant of the crystal is denoted by 
$a_1=a_2=a_3=a$. For the electron-ion interaction $V_{ion}(\vec{r}\,)$, 
Eq. (\ref{Vk_3dPot}) of the Doyle-Turner model is used in the calculation. All figures 
for the calculated $V_{2D}(x,y)$ are plotted with germanium (Ge) lattice and the fitting 
parameters in the Doyle-Turner model obtained by Chouffani and \"{U}berall \cite{Chouf99}. 
For the convenience of reading, those fitting parameters are re-listed in Table II. 
The value used for the root-mean-square amplitude of the lattice thermal vibration of Ge
in Eq. (\ref{Vk_3dPot}) is $\sqrt{\left<\mu^2\right>}=0.085$\AA\, at a temperature of 
293 K \cite{Gemme74}. The lattice constant of Ge is $a=5.658$\AA.

\vspace{0.1in}
\noindent
{\bf a. Channeling along [001] Crystal Axis}
\vspace{0.1in}

For the channeling along the [001] crystal axis, the beam longitudinal coordinate $z$ is 
aligned with the lattice coordinate $X_3$. A simple choice of the transverse coordinate 
is $x$ and $y$ aligned with [100] and [010] crystal direction, respectively, and the 
beam coordinate $(x,y,z)$ is aligned with the lattice coordinate $(X_1,X_2,X_3)$. 
In Eq. (\ref{Rotation}), on the other hand, $\theta_1=0$ and $\theta_2=\pi/2$ for 
$[hkl]=[001]$, which yields a transformation matrix ${\bf R}$ that switches only the 
$x$ and $y$ coordinate and has no any physical consequence on the Fourier transformation 
of the lattice potential. No coordinate transformation is therefore needed, $i.e.$ 
${\bf R}$ can be chosen as an identity matrix and the lattice potential $V_{2D}(x,y)$ 
can easily be calculated from Eqs. (\ref{Vk_3dPot}) and (\ref{V2dFourier}). In the 
contour plot of the calculated $V_{2D}(x,y)$, however, the axes of the primitive unit 
cell of the two-dimensional lattice on the transverse plane were observed to be aligned 
with the [110] and $[\bar{1}10]$ crystal direction. A rotational transformation in the 
transverse plane is therefore needed to align $x$ and the $y$ axes with the [110] and 
$[\bar{1}10]$ crystal direction, respectively, $i.e.$ 
\begin{equation}
\left(\begin{array}{l} X_1 \vspace{0.05in}\\ X_2 \vspace{0.05in}\\ X_3 \end{array}\right)
=\left(\begin{array}{ccc} 
 \cos(\pi/4) & -\sin(\pi/4) & \hspace{0.1in} 0 \hspace{0.1in}  \vspace{0.05in}\\ 
 \sin(\pi/4) & \;\;\;\cos(\pi/4)  & 0 \vspace{0.05in}\\ 
   0         & \;\;\; 0     & 1          
\end{array}\right) 
\left(\begin{array}{l} x \vspace{0.05in}\\ y \vspace{0.05in}\\ z \end{array}\right)
={\bf R}^{-1}
\left(\begin{array}{l} x \vspace{0.05in}\\ y \vspace{0.05in}\\ z \end{array}\right)
\end{equation}
where $\pi/4$ is the rotation angle from the [100] and [010] direction to the [110] and 
$[\bar{1}10]$ direction. From Eq. (\ref{Grot}), the transformation of the reciprocal 
lattice vector is
\begin{equation}
\label{GtoG001}
\left(\,\frac{k_1+k_2}{\sqrt{2}\,a}\;,\; 
    \frac{k_2-k_1}{\sqrt{2}\,a}\;,\;
    \frac{k_3}{a} \,\right) 
=
\left(\,\frac{n_1}{b_1}\;,\;\frac{n_2}{b_2}\;,\;\frac{n_3}{b_3}\,\right)
\end{equation}
Since for each given set of integers $(n_1,n_2,n_3)$, $(k_1,k_2,k_3)$ have to be solved 
from Eq. (\ref{GtoG001}) as integers, the solution with the minimal values of 
$(b_1,b_2,b_3)$ is 
\begin{equation}
\left\{\begin{array}{lll}
\left(b_1\,,\,b_2\,,\,b_3 \right)&=&
              \left(\sqrt{2}\,a/2\,,\,\sqrt{2}\,a/2\,,\,a \right) \vspace{0.05in} \\
\left(k_1\,,\,k_2\,,\,k_3 \right)&=&
              \left(n_1-n_2\,,\,n_1+n_2\,,\,n_3 \right) \vspace{0.01in}
\end{array}\right.
\end{equation}
The lattice potential with $x$ and $y$ coordinate aligned with the $[110]$ and 
$[\bar{1}10]$ crystal axes is then calculated from Eqs. (\ref{V(xy)}), 
(\ref{V2dFourier}), and (\ref{Vk_3dPot}) as
\begin{equation}
\label{V001}
V_{2D}(x,y) = \sum_{n_1=-\infty}^{\infty}\sum_{n_2=-\infty}^{\infty} V_{n_1n_2}
        \exp\left[i2\pi\left(\frac{n_1x}{a/\sqrt{2}}+\frac{n_2y}{a/\sqrt{2}}\right)\right]
\end{equation}
where
\begin{eqnarray}
\label{V100Vk}
V_{n_1n_2} 
&=&
  -\frac{2\pi \hbar^2}{m_ev_0} \sum_{\vec{k}} \left(
   \sum_{i=1}^{M}\alpha_i\,e^{-\lambda_i^2|\vec{k}|^2/(2a)^2} \;
   \sum_{j=1}^{N} \,e^{-i2\pi(\vec{k}\cdot\vec{r}_j)/a} \right) 
   \delta_{k_1,(n_1-n_2)}\delta_{k_2,(n_1+n_2)}\delta_{k_3,0}
   \nonumber \\
&=& -\frac{2\pi \hbar^2}{m_ea^2}
 \sum_{i=1}^{M}\frac{\alpha_i}{a}\, e^{-\lambda_i^2(n_1^2+n_2^2)/2a^2}
 \sum_{j=1}^{N} e^{-i2\pi[n_1(x_j+y_j)-n_2(x_j-y_j)]/a} 
\end{eqnarray}
With the coordinate $\vec{r}_j$ of the ions in a unit cell given in Table I, the second 
summation in Eq. (\ref{V100Vk}) can be evaluated as
\begin{equation}
\sum_{j=1}^{N} e^{-i2\pi[n_1(x_j+y_j)-n_2(x_j-y_j)]/a} = \left\{
\begin{array}{lll}
8\,, &\hspace{0.2in} &\mbox{for $(n_1,n_2)=\mbox{(even, even)}$} \vspace{0.05in}\\
0\,, & & \mbox{otherwise}
\end{array}\right.
\end{equation} 
To purge the terms of zero in the summation of Eq. (\ref{V001}), let $n_1=2k_1$ and 
$n_2=2k_2$, and the lattice potential in the transverse plane can be rewritten as
\begin{equation}
\label{V001_2}
V_{2D}(x,y) = \sum_{k_1=-\infty}^{\infty}\sum_{k_2=-\infty}^{\infty} v_{k_1k_2}
         \exp\left[i\frac{2\pi}{a/(2\sqrt{2})}\left(k_1x+k_2y\right)\right] 
\end{equation}
where
\begin{equation}
\label{Vk001_2}
v_{k_1k_2} = V_{(2k_1)(2k_2)} = -\frac{16\pi \hbar^2}{m_ea^2}
            \sum_{i=1}^{M}\frac{\alpha_i}{a}\, e^{-2\lambda_i^2(k_1^2+k_2^2)/a^2}
\end{equation}
The periods of $V_{2D}(x,y)$ is therefore $a_x=a_y=a/(2\sqrt{2})$ in both $x$ and $y$ 
direction when $x$ and $y$ axes are aligned with the $[110]$ and $[\bar{1}10]$ crystal 
axis that are of the primitive cell in the transverse plane of a beam channeling in the 
[001] direction. The lattice potential in Eqs. (\ref{V001_2}) and (\ref{Vk001_2}) has 
also been obtained previously using the method of strings of ions in the axial channeling 
model \cite{Chouf99}, as this is the easiest case for the axial channeling calculation. 
As shown in Fig. 2a, $V_{2D}(x,y)$ has a single potential well in each square unit cell 
on the beam transverse plane. Near the bottom of the potential wells, where the strings 
of ions is located, the interaction is almost rotationally symmetric on the transverse 
plane. The rotational symmetry of the lattice potential has been previously used to 
simplifying the study of the Bloch eigenstates and transitions between the eigenstates 
for the transverse motion of beam electrons \cite{Ander82,Klein,Chouf99,Genz96}.

\vspace{0.1in}
\noindent
{\bf b. Channeling along [110] Crystal Axis}
\vspace{0.1in}

For a beam channeling in $[hkl]=[110]$ direction, $\theta_1=\arctan(k/h)=\pi/4$ and 
$\theta_2=0$ for transformation matrix ${\bf R}$ in Eq. (\ref{Rotation}) and $h_y=k_y=0$ 
and $l_y=2$ for the Miller indices of the crystal directions on the transverse plane. 
The $x$ and $y$ coordinate on the transverse plane are therefore aligned with 
$[\bar{k}hl]=[\bar{1}10]$ and $[h_yk_yl_y]=[001]$ crystal axes, respectively. The 
transformation of the reciprocal lattice vector is calculated from Eq. (\ref{Grot}) as
\begin{equation}
\label{GtoG110}
\left(\,\frac{k_2-k_1}{\sqrt{2}\,a}\;,\; 
        \frac{k_3}{a}\;,\;
        \frac{k_1+k_2}{\sqrt{2}\,a} \,\right) 
=  \left(\,\frac{n_1}{b_1}\;,\;\frac{n_2}{b_2}\;,\;\frac{n_3}{b_3}\,\right)
\end{equation}
For integer $(k_1,\,k_2,\,k_3)$ and $(n_1,\,n_2,\,n_3)$, the solution of Eq. (\ref{GtoG110})
with the minimal values of $(b_1,b_2,b_3)$ is 
\begin{equation}
\label{110kTOk}
\left\{\begin{array}{lll}
\left(b_1\,,\,b_2\,,\,b_3 \right)&=&\left(\sqrt{2}\,a/2\,,\,a\,,\,\sqrt{2}\,a/2\,\right) 
     \vspace{0.05in} \\
\left(k_1\,,\,k_2\,,\,k_3 \right)&=&\left(n_3-n_1\,,\,n_1+n_3\,,\,n_2\right)\vspace{0.01in}
\end{array}\right.
\end{equation}
The lattice potential with $x$ and $y$ coordinate aligned with the $[\bar{1}10]$ and [001] 
crystal axes is then calculated from Eqs. (\ref{V(xy)}), (\ref{V2dFourier}), and 
(\ref{Vk_3dPot}) as
\begin{equation}
\label{V110}
V_{2D}(x,y) = \sum_{n_1=-\infty}^{\infty}\sum_{n_2=-\infty}^{\infty} V_{n_1n_2}
              \exp\left[i2\pi\left(\frac{n_1x}{a/\sqrt{2}}+\frac{n_2y}{a}\right)\right] 
\end{equation}
and
\begin{equation}
\label{V110Vk}
V_{n_1n_2} = -\frac{2\pi \hbar^2}{m_ea^2}
 \sum_{i=1}^{M}\frac{\alpha_i}{a}\, e^{-\lambda_i^2(2n_1^2+n_2^2)/(2a)^2} \,
 \sum_{j=1}^{N} e^{-i2\pi[n_1(y_j-x_j)+n_2z_j]/a}
\end{equation}
Since there is no additional periodic zero in the summation over $(x_j,y_j,z_j)$ in 
Eq. (\ref{V110Vk}), the expansion in Eq. (\ref{V110}) is the correct Fourier expansion 
of $V_{2D}(x,y)$ for solving the Bloch eigenstates. As shown in Fig. 2b, there are four
potential wells in each unit cell and the lattice potential is highly anisotropic in the 
transverse plane. The period of $V_{2D}(x,y)$ are $a_x=a/\sqrt{2}$ and $a_y=a$ along 
the $[\bar{1}10]$ and $[001]$ crystal axes that are the axes of the primitive unit cell 
in the transverse plane for a beam channeling through a diamond-like crystal along the 
$[110]$ crystal axis.

\vspace{0.1in}
\noindent
{\bf c. Channeling along [111] Crystal Axis}
\vspace{0.1in}

For a beam channeling in $[hkl]=[111]$ direction, $\theta_1=\arctan(k/h)=\pi/4$ and 
$\theta_2=\arctan(l/\sqrt{h^2+k^2})=\arctan(1/\sqrt{2})$ for transformation matrix 
${\bf R}$ in Eq. (\ref{Rotation}) and $h_y=-hl=-1$, $k_y=-kl=-1$, and $l_y=h^2+k^2=2$ 
for the Miller indices of one crystal direction for the transverse coordinates. 
The $x$ and $y$ coordinate are therefore aligned with $[\bar{k}h0]=[\bar{1}10]$ and 
$[h_yk_yl_y]=[\bar{1}\bar{1}2]$ crystal axes, respectively. The transformation of 
the reciprocal lattice vector is calculated from Eq. (\ref{Grot}) as
\begin{equation}
\label{TranG111}
\left(\,\frac{k_2-k_1}{\sqrt{2}\,a} \;,\;
        \frac{2k_3-k_1-k_2}{\sqrt{6}\,a} \;,\; 
        \frac{k_1+k_2+k_3}{\sqrt{3}\,a} \, \right) 
= \left(\,\frac{n_1}{b_1}\;,\;\frac{n_2}{b_2}\;,\;\frac{n_3}{b_3}\,\right)
\end{equation}
For integer $(k_1,\,k_2,\,k_3)$ and $(n_1,\,n_2,\,n_3)$, the solution of Eq. (\ref{TranG111})
with the minimal values of $(b_1,b_2,b_3)$ at $n_3=0$ is
\begin{equation}
\label{Solu111}
\left\{\begin{array}{lll}
\left(b_1\,,\,b_2\,,\,b_3 \right)
   &=& \left(\,a/\sqrt{2}\;,\;a/\sqrt{6}\;,\;\sqrt{3}a \,\right)  \vspace{0.05in} \\
\left(k_1\,,\,k_2\,,\,k_3 \right)
   &=& \left(-n_1-n_2\;,\;n_1-n_2\;,\;2n_2 \,\right)\vspace{0.01in}
\end{array}\right.
\end{equation}
The 2D lattice potential can then be calculated from Eqs. (\ref{V(xy)}), (\ref{V2dFourier}), 
and (\ref{Vk_3dPot}) as
\begin{equation}
\label{V111}
V_{2D}(x,y) = \sum_{n_1=-\infty}^{\infty}\sum_{n_2=-\infty}^{\infty} V_{n_1n_2}
     \exp\left[i2\pi\left(\frac{n_1x}{a/\sqrt{2}}+\frac{n_2y}{a/\sqrt{6}}\right)\right] 
\end{equation}
where
\begin{equation}
\label{V111Vk}
V_{n_1n_2} = -\frac{2\pi \hbar^2}{m_e a^2}
 \sum_{i=1}^{M}\frac{\alpha_i}{a}\, e^{-\lambda_i^2(n_1^2+3n_2^2)/2a^2} \,
 \sum_{j=1}^{N} e^{i2\pi[n_1(x_j-y_j)+n_2(x_j+y_j-2z_j)]/a}
\end{equation}
With the ion coordinates $(x_j,y_j,z_j)$ given in Table I, the second summation in 
Eq. (\ref{V111Vk}) can be evaluated as
\begin{equation}
\sum_{j=1}^{N} e^{i2\pi[n_1(x_j-y_j)+n_2(x_j+y_j-2z_j)]/a} = \left\{
\begin{array}{lll}
8\,, &\hspace{0.2in} &\mbox{for $n_1+n_2=\mbox{even}$} \vspace{0.05in}\\
0\,, & & \mbox{otherwise}
\end{array}\right.
\end{equation} 
and $V_{n_1n_2}$ can thus be rewritten as
\begin{equation}
V_{n_1n_2} = -\frac{16\pi \hbar^2}{m_e a^2}
 \left(\sum_{i=1}^{M}\frac{\alpha_i}{a}\, e^{-\lambda_i^2(n_1^2+3n_2^2)/2a^2} \right)
 \delta_{n_1+n_1,\mbox{even}}
\end{equation}
Because of different periodicity of $V_{2D}(x,y)$ in Eq. (\ref{V111}) along the $x$ and 
$y$ direction, the periodic occurrence of zero in $V_{n_1n_2}$ cannot be removed from 
the summations in Eq. (\ref{V111}) by relabeling the summation indices and a rotation of 
$(x,y)$ coordinate while keeping the form of a Fourier expansion. This can also be seen 
in the contour plot of $V_{2D}(x,y)$ in Fig. 2c, where there are two identical potential 
wells at the corners and center of a unit cell and it cannot be simplified into a single 
square lattice because of different lattice constants along the $x$ and $y$ direction. The 
$[\bar{1}10]$ and $[\bar{1}\bar{1}2]$ crystal axes are therefore the orthogonal axes of 
the primitive unit cell in the transverse plane and $a_x=a/\sqrt{2}$ and $a_y=a/\sqrt{6}$ 
are the correct periods of $V_{2D}(x,y)$ for a beam channeling in the $[111]$ direction. 

\vspace{0.1in}
\noindent
{\bf d. Channeling along [210] Crystal Axis}
\vspace{0.1in}

For a beam channeling along the $[hkl]=[210]$ crystal axis, 
$\theta_1=\arctan(k/h)=\arctan(1/2)$ and $\theta_2=\arctan(l/\sqrt{h^2+k^2})=0$ and 
the solution of Eq. (\ref{Grot}) is
\begin{equation}
\label{Solu210}
\left\{\begin{array}{lll}
\left(b_1\,,\,b_2\,,\,b_3 \right)
   &=& \left(\,a/\sqrt{5}\;,\; a \,,\; a/\sqrt{5} \;\right)  \vspace{0.05in} \\
\left(k_1\,,\,k_2\,,\,k_3 \right)
   &=& \left(-n_1+2n_3\;,\;2n_1+n_3\;,\;n_2 \,\right)\vspace{0.01in}
\end{array}\right.
\end{equation}
The lattice potential with $x$ and $y$ coordinate aligned with $[\bar{k}hl]=[\bar{1}20]$ 
and $[h_yk_yl_y]=[001]$ crystal axes can be calculated from Eqs. (\ref{V(xy)}), 
(\ref{V2dFourier}), and (\ref{Vk_3dPot}) as
\begin{equation}
\label{V210}
V_{2D}(x,y) = \sum_{n_1=-\infty}^{\infty}\sum_{n_2=-\infty}^{\infty} V_{n_1n_2}
              \exp\left[i2\pi\left(\frac{n_1x}{a/\sqrt{5}}+\frac{n_2y}{a}\right)\right] 
\end{equation}
where
\begin{equation}
\label{V210Vk}
V_{n_1n_2} = -\frac{2\pi \hbar^2}{m_e a^2}
 \sum_{i=1}^{M}\frac{\alpha_i}{a}\, e^{-\lambda_i^2(5n_1^2+n_2^2)/(2a)^2} \,
 \sum_{j=1}^{N} e^{-i2\pi[n_1(2y_j-x_j)+n_2z_j]/a}
\end{equation}
With ion coordinates $\vec{r}_j$ in a unit cell given in Table I, the summation over the 
ions in Eq. (\ref{V210Vk}) is
\begin{equation}
\sum_{j=1}^{N} e^{-i2\pi[n_1(2y_j-x_j)+n_2z_j]/a} = \left\{
\begin{array}{lll}
{\displaystyle 4\left[1+(-1)^{(n_1+n_2)/2} \right]} \,, 
     &\hspace{0.05in} &\mbox{for $(n_1,n_2)=\mbox{(even, even)}$} \vspace{0.05in}\\
0\,, & & \mbox{otherwise}
\end{array}\right.
\end{equation} 
Similar to the case of the channeling in the [001] direction, this periodic occurrence of 
zero in the Fourier expansion can be removed by relabelling the summation indices as
$n_1=2k_1$ and $n_2=2k_2$ in Eq. (\ref{V210}) without altering the form of the Fourier 
expansion. The periods of $V_{2D}(x,y)$ in this case is thus $a_x=a/(2\sqrt{5})$ and 
$a_y=a/2$ along the $[\bar{1}20]$ and $[001]$ crystal axes, respectively, and the Fourier 
expansion of the 2D lattice potential in Eq. (\ref{V210}) should be rewritten as
\begin{equation}
V_{2D}(x,y) = \sum_{k_1=-\infty}^{\infty}\sum_{k_2=-\infty}^{\infty} v_{k_1k_2}
              \exp\left[i2\pi\left(\frac{k_1x}{a/(2\sqrt{5})}+\frac{k_2y}{a/2}\right)\right] 
\end{equation}
where
\begin{equation}
v_{k_1k_2} = -\frac{8\pi \hbar^2}{m_e a^2}
 \sum_{i=1}^{M}\frac{\alpha_i}{a}\, e^{-\lambda_i^2(5k_1^2+k_2^2)/a^2} 
 \left[1+(-1)^{k_1+k_2}\right]
\end{equation}
where $v_{k_1k_2}=0$ for $k_1+k_2=odd$. As shown in Fig. 2d, there are two identical 
potential wells in each unit cell of $V_{2D}(x,y)$. Similar to the case of channeling in 
the [111] direction, this 2D body-center rectangular unit cell cannot be simplified into 
a simple rectangular unit cell because of different periodicity of $V_{2d}(x,y)$ along the 
$x$ and $y$ direction. The $[\bar{1}20]$ and $[001]$ crystal axis are the orthogonal axes 
of the primitive unit cell in the transverse plane and $a_x=a/(2\sqrt{5})$ and $a_y=a/2$ 
are the correct periodicity of $V_{2D}(x,y)$ for a beam channeling through a diamond-like 
crystal along the $[210]$ crystal axis.

\section{Final Remarks}

To solve the Bloch eigenstates for the transverse motion of beam electrons in a crystal 
numerically, the Fourier expansion of the lattice potential in Eq. (\ref{V(xy)}) needs 
to be truncated as 
\begin{equation}
\label{truncated}
V_{2D}(x,y)=\sum_{n_1=-K_{max}}^{K_{max}} \sum_{n_2=-K_{max}}^{K_{max}} 
            V_{n_1n_2}\,e^{i2\pi(n_1x/b_1+n_2y/b_2)}
\end{equation} 
This truncation is possible due to a fast decay of expansion coefficient $V_{n_1n_2}$
which is the consequence of an electron-ion interaction potential that decays faster 
then $1/r$. The convergence of the truncation, however, needs to be checked to ensure the 
accuracy of the numerically calculated potential. To examine the convergence, we estimate 
the truncation error using 
\begin{equation}
\label{Error}
 \mbox{Truncation Error} = \frac{1}{a_xa_y}\int\limits_0^{a_y}\int\limits_0^{a_x}
    \left|\frac{V_{2D}(x,y,K_{max})}{V_{2D}(x,y,K_{max}+\Delta)}-1\right|dxdy
\end{equation}
where $V_{2D}(x,y,K_{max})$ and $V_{2D}(x,y,K_{max}+\Delta)$ are the lattice potential 
$V_{2D}(x,y)$ calculated with the truncation at $K_{max}$ and $K_{max}+\Delta$, 
respectively. Figure 3 plots this truncation error as a function of $K_{max}$ for the 
examples in Fig. 2 and shows that the numerically calculated potential is sufficiently 
accurate (with a truncation error smaller than $10^{-8}$) for $K_{max}<50$. Note that 
solving the Bloch eigenstates numerically from a two-dimensional Schr\"{o}dinger equation 
requires diagonalizations of $(2K_{max}+1)^2\times(2K_{max}+1)^2$ matrices for the 
Hamiltonian operator and, with $K_{max}<50$, this computational task can be handled with 
well-configured pc computers nowadays.

In summary, we have developed a general formalism for the lattice potential for studying 
the transverse motion of beam electrons when an ultra-relativistic beam channels through 
a crystal lattice. As shown by the examples, this two-dimensional lattice potential can 
easily be calculated for a beam channeling through a crystal along any crystal direction. 
With the availability of the 2D lattice potential and increasing power of pc computers, 
the 2D calculation of the Bloch eigenstates should becomes the standard for the study of 
the channeling radiation and the 1D approximation in modeling the channeling radiation is 
no longer needed.

\newpage

\begin{center}
\begin{minipage}{6.2in}
Table I. Ion position $\vec{r}_j$ in a unit cell of diamond-like crystals where the origin 
of the coordinate is at a corner of the unit cell and $a$ is the lattice constant.

\begin{center}\begin{tabular}{ccccccccc} \hline\hline
 $j$ 
 & \hspace{0.05in}1\hspace{0.05in} 
 & \hspace{0.15in} 2\hspace{0.15in} 
 & \hspace{0.15in}3\hspace{0.15in} 
 & \hspace{0.15in}4\hspace{0.15in} 
 & \hspace{0.15in}5\hspace{0.15in} 
 & \hspace{0.15in}6\hspace{0.15in} 
 & \hspace{0.15in}7\hspace{0.15in} 
 & \hspace{0.15in}8\hspace{0.15in}         \\ \hline
 \;\;$x_j/a$\;\; & 0 & 1/4 &   0 & 1/4 & 1/2 & 3/4 & 1/2 & 3/4 \\ \hline
 $y_j/a$ & 0 & 1/4 & 1/2 & 3/4 &   0 & 1/4 & 1/2 & 3/4 \\ \hline
 $z_j/a$ & 0 & 1/4 & 1/2 & 3/4 & 1/2 & 3/4 &  0  & 1/4 \\ \hline
\end{tabular}\end{center} 
\end{minipage}

\vspace{0.5in}
\begin{minipage}{6.2in}
Table II. Fitting parameters $\alpha_j$ and $\beta_j$ for atomic form factor of germanium 
obtained by Chouffani and \"{U}berall \cite{Chouf99}, where $a=5.658$\AA\, is the lattice 
constant. 

\begin{center}\begin{tabular}{ccccccc}\hline\hline
      $i$ & 1 & 2 & 3 & 4 & 5 & 6 \\\hline
\hspace{0.05in} $\alpha_i/a$ \hspace{0.05in} &
 \; 0.3204585\; & \; 0.4895285\; & \; 0.1625405\; & \; 0.0121414\; & 
 \; 0.0427155\; & \; 0.0027543\; \\\hline
 $\beta_i/a^2$ &   
 \; 2.0642597\, & \; 0.6336834\, & \; 0.0351070\, & \; 0.0010696\, & 
 \; 0.0073946\, & \; 0.1409283\, \\\hline
\end{tabular}\end{center}
\end{minipage}
\end{center}

\newpage

\noindent
\begin{minipage}{6.5in}

\vspace{-0.5in}
\hspace{-1.7in}
\begin{minipage}{3in}
\includegraphics[scale=1.2]{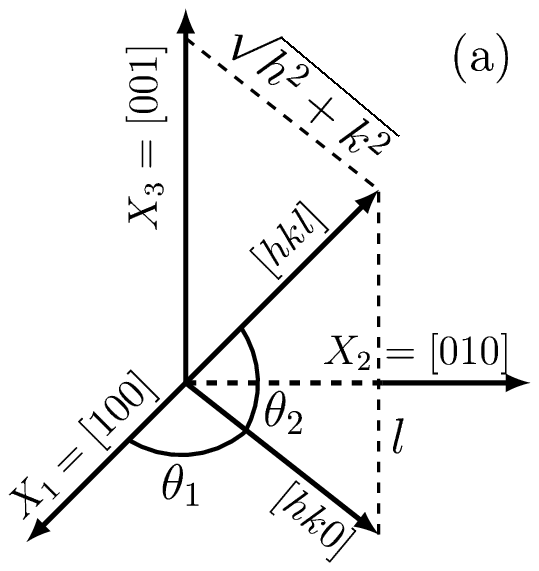}
\end{minipage}\hspace{0.2in}
\begin{minipage}{3in}
\includegraphics[scale=1.2]{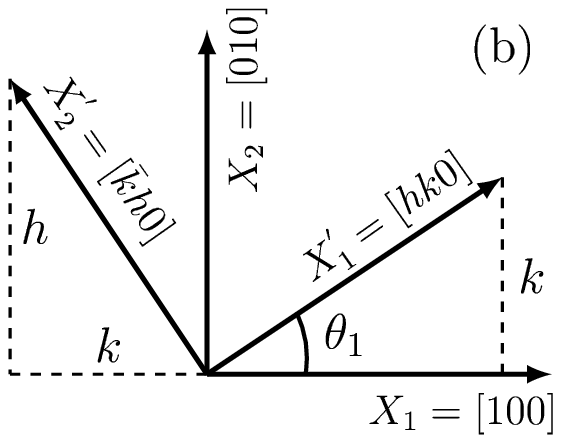}
\end{minipage} 

\vspace{-10in}
\begin{minipage}{3in}
\includegraphics[scale=1.2]{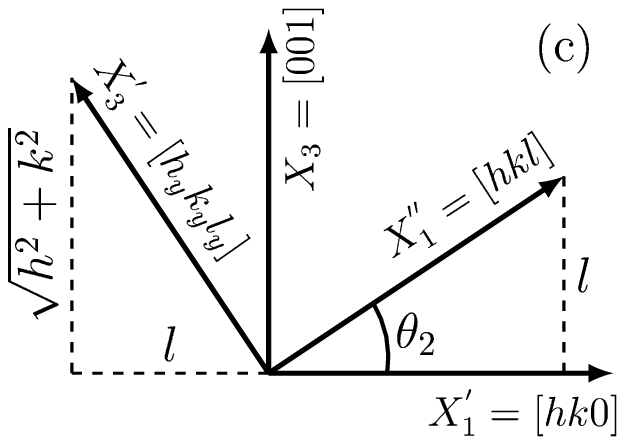}
\end{minipage} 

\vspace{-9.3in}
\noindent
Figure 1. (a) The $[hkl]$ and $[hk0]$ crystal axis in the lattice coordinate 
$(X_1,X_2,X_3)$ of an orthorhombic crystal lattice, (b) the 1st coordinate rotation on 
the $X_1$--$X_2$ plane and (c) the 2nd coordinate rotation on the $X'_1$--$X_3$ plane 
for the construction of ${\bf R}$ for a beam channeling in $[hkl]$ direction, where 
$\theta_1=\arctan(k/h)$ and $\theta_2=\arctan(l/\sqrt{h^2+k^2})$. The crystal axis 
$[h_yk_yl_y]$ is along the direction of integer vector 
$(h_y,k_y,l_y)=(h,k,l)\times(-k,h,0)=(-hl,-kl,h^2+k^2)$ 
\end{minipage}
  
\newpage
\noindent
\begin{minipage}{3.2in}
\epsfig{file=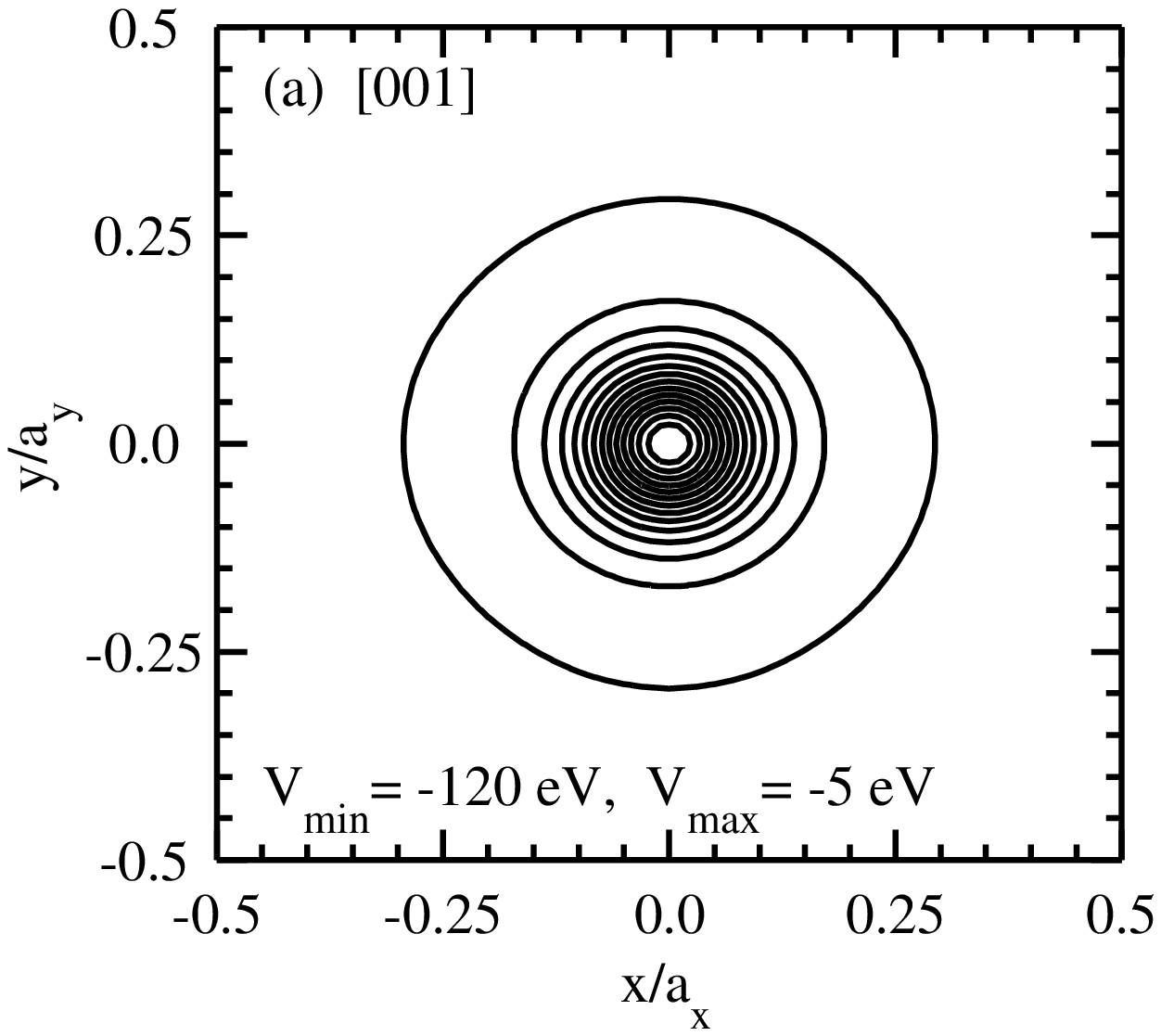, width=80mm, angle=-90} 
\end{minipage}\hspace{0.1in} 
\begin{minipage}{3.2in}
\epsfig{file=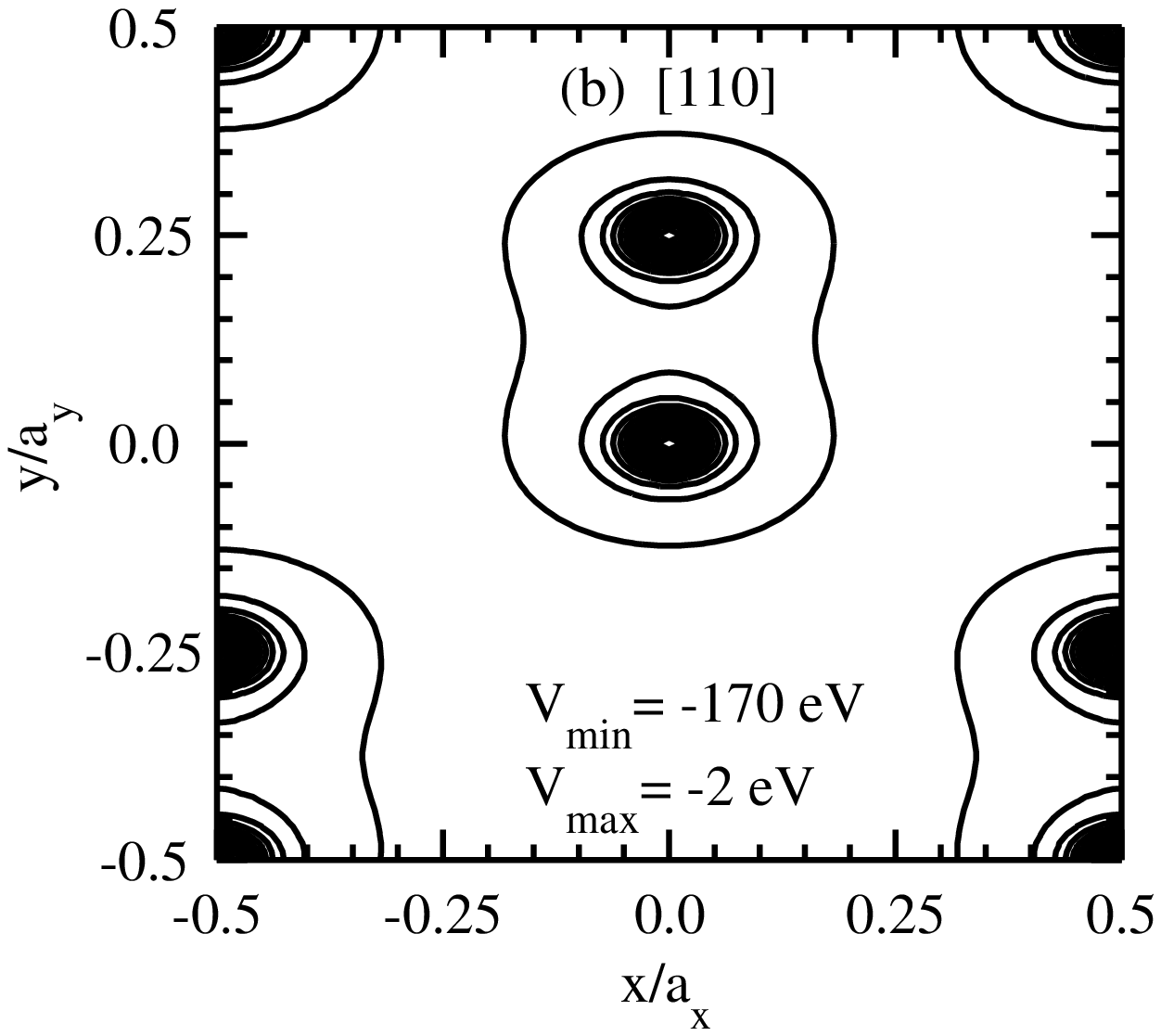, width=80mm, angle=-90} 
\end{minipage} 

\noindent
\begin{minipage}{3.2in}
\epsfig{file=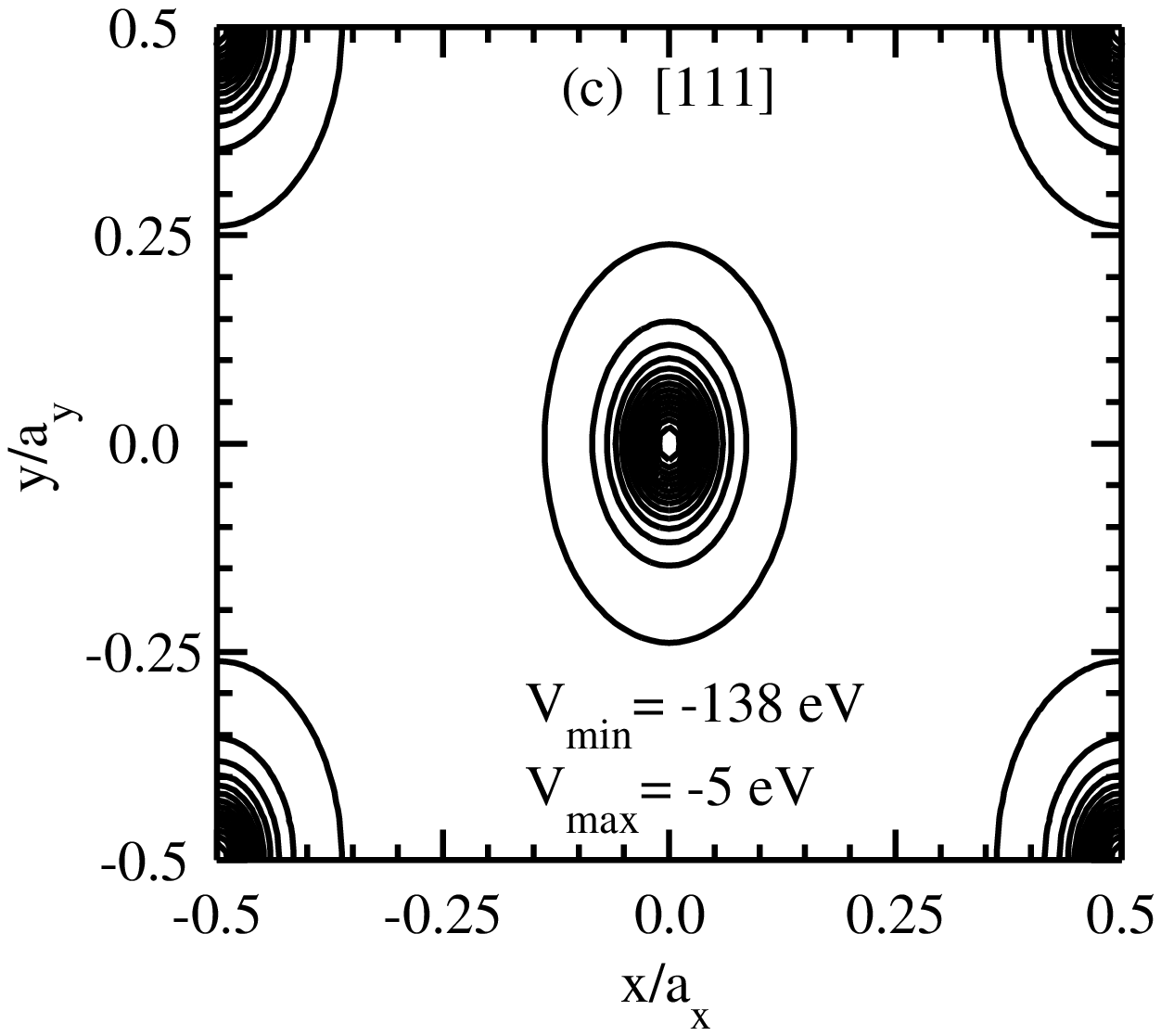, width=80mm, angle=-90} 
\end{minipage}\hspace{0.1in} 
\begin{minipage}{3.2in}
\epsfig{file=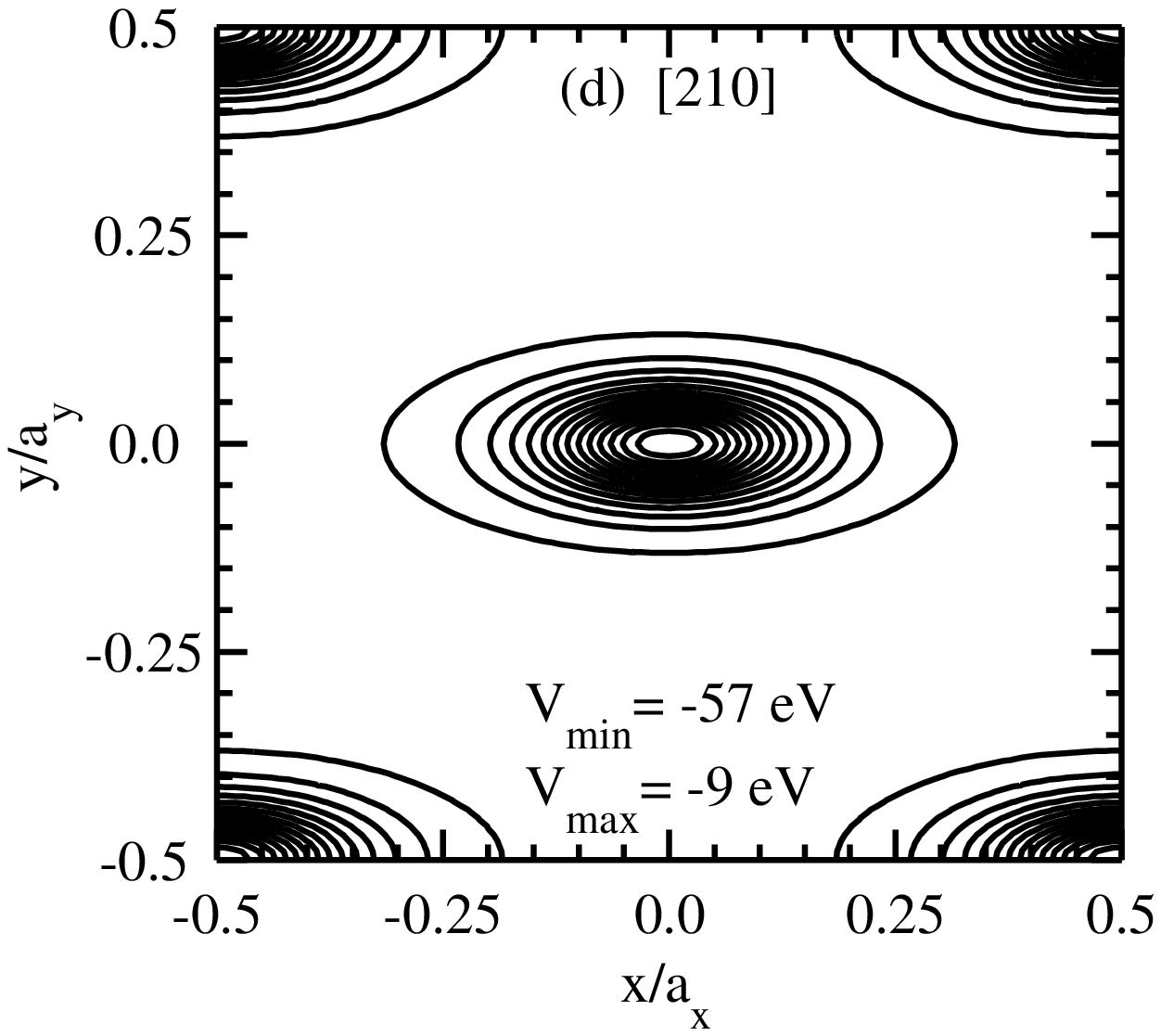, width=80mm, angle=-90} 
\end{minipage} 

\vspace{0.1in}
\noindent
Figure 2. Contour plots of lattice potential $V_{2D}(x,y)$ on the beam transverse plane 
for a beam channeling through a diamond-like crystal along (a) [001], (b) [110], (c) [111], 
and (d) [210] direction, where $a_x$ and $a_y$ are the periods of $V_{2D}(x,y)$ in the $x$ 
and $y$ direction, respectively, and $V_{min}$ and $V_{max}$ are the minimum and maximum
of $V_{2D}(x,y)$. The centers of equipotential loops are the bottoms of potential wells.

\newpage
\begin{center}
\epsfig{file=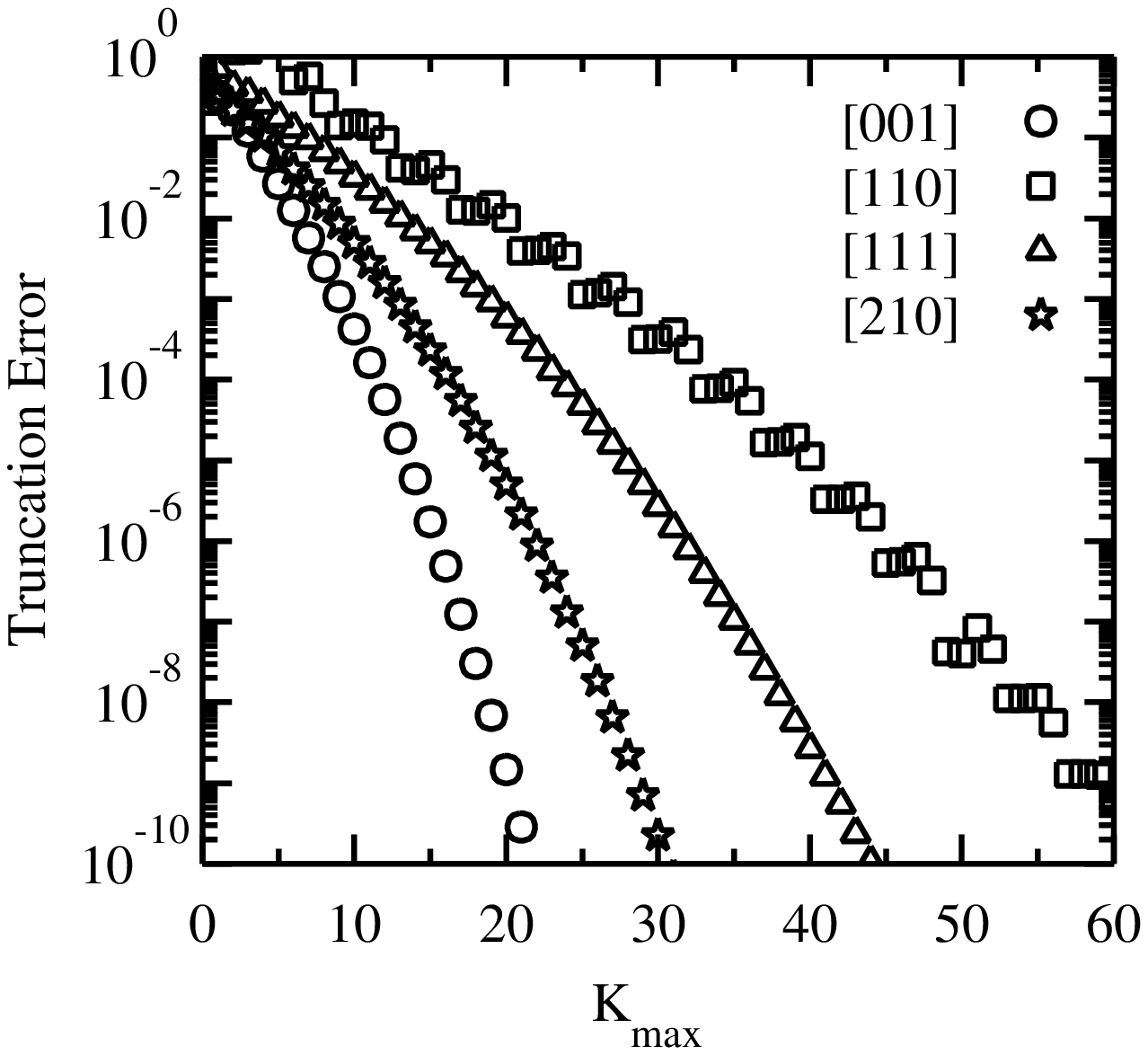, width=120mm, angle=-90} 
\end{center}

\vspace{0.1in}
\noindent
Figure 3. Truncation error of the truncated Fourier expansion of $V_{2D}(x,y)$ in 
Eq. (\ref{truncated}) $v.s.$ $K_{max}$ calculated using Eq. (\ref{Error}) with $\Delta=5$ 
for channeling of germanium along the [001] (circles), [110] (squares), [111] (triangles),
and [210] (stars) crystal axis, respectively. 
  
\end{document}